\documentclass[%
 reprint,
 amsmath,amssymb,
 aps,
pra,
]{revtex4-2}

\usepackage{graphicx}
\usepackage{dcolumn}
\usepackage{bm}
\usepackage{bbm}
\usepackage{float}
\usepackage{amsmath}
\usepackage{mathtools}
\usepackage{url}
\usepackage[normalem]{ulem}
\usepackage{hyperref}
\hypersetup{
   unicode=true,
   plainpages=false,
   colorlinks=true,
   citecolor=blue,
   urlcolor=blue
}
\usepackage[caption=false,position=top,singlelinecheck=off,justification=raggedright]{subfig}
\allowdisplaybreaks

\begin{document}

\preprint{APS/123-QED}

\title{Bosonization in $R$-paraparticle Luttinger models}

\author{Dennis F. Salinel}
    \email{dfsalinel@up.edu.ph}
    \affiliation{National Institute of Physics, University of the Philippines Diliman, Quezon City 1101, Philippines}
    
\author{Kristian Hauser A. Villegas}
    \email{kavillegas1@up.edu.ph}
    \affiliation{National Institute of Physics, University of the Philippines Diliman, Quezon City 1101, Philippines}

\date{\today}

\begin{abstract}
Alternative theories of quantum statistics provide an avenue for exploring novel physics beyond bosons and fermions, yet experimental verification of their existence in nature proves a challenging task. Among these theories, it has recently been suggested that $R$ parastatistics can be realized as quasiparticle excitations in many-body systems. In this paper, we build on this idea by showing that signatures of $R$ parastatistics can be observed as flavor-charge separation in one-dimensional (1D) systems. We consider a generalized version of the Luttinger model (LM) and show that bosonization persists when the $R$ paraparticles have Fermi-surface-like structures. These $R$ \textit{parafermions} can satisfy generalized exclusion principles beyond conventional Pauli's. We show that density waves of all $R$ parafermions can always be bosonized, but flavor waves act like bosons only for a certain subclass of $R$ parafermions. We derive the conditions for bosonization by analyzing the LM spectrum, showing that bosonization applies only to low-temperature systems. Signatures of flavor-charge separation then become apparent as distinct dispersion profiles when we turn on interparticle interactions. This points to potential observations of flavor-charge separation in 1D systems that host emergent $R$ paraparticles.
\end{abstract}

\maketitle


\section{Introduction}\label{sec:intro}

Conventionally, particles are categorized as either bosons or fermions; their difference lies primarily with their inherent exchange statistics, where the boson (fermion) wavefunction is symmetric (antisymmetric) under particle exchange. One popular exception to this dichotomy is anyons \cite{Wilczek_1982_1, Wilczek_1982_2}, existing only in two dimensions \cite{Nakamura_2020, Bartolomei_2020} and where the particle's wavefunction picks up an arbitrary phase under an exchange of particles.

\begin{figure*}[t]
    \centering
    \includegraphics[width=\linewidth]{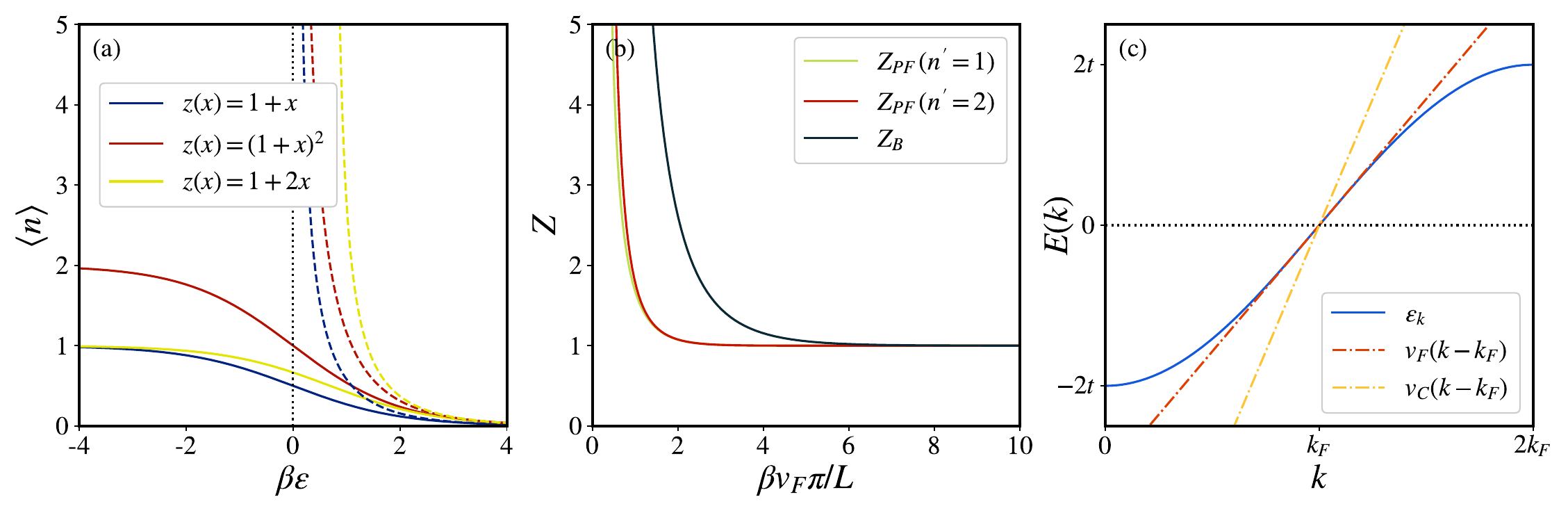}
    \caption{Summary of main results. (a.) Occupation statistics of different types of $R$ parafermions (solid line) and $R$ parabosons (dashed line). The single-mode partition functions of the latter are obtained using Eq.~\eqref{eq:a1}. (b) Partition functions of the noninteracting $R$-paraparticle LMs in the $R$-parafermion and boson basis. (c) Sketch of expected dispersions of $R$-parafermionic quasiparticles obtained from an underlying spinor TG gas.}
    \label{fig:1}
\end{figure*}

Outside of anyons, multiple theories of unconventional quantum statistics have also been proposed \cite{Green_1953, Haldane_1991, Greenberg_1990, Greenberg_1991}, and although they present novel physical predictions \cite{Takahiro_1995, Kazumoto_1998, Greenberg_1999, Ebadi_2013}, no unambiguous experimental evidence of these unconventional particles has yet been reported. Among these theories is Green's parastatistics \cite{Green_1953}, which remains a subject of ongoing discussion concerning its actual occurrence in nature \cite{mekonnen_2025, Toppan_2025}. While superselection rules \cite{Greenberg_1965} and no-go theorems \cite{DHR_I, DHR_II} tell us that under certain assumptions Green's paraparticles can always be reexpressed in terms of ordinary bosons and fermions \cite{Araki_1961, Druhl_1970}, they do not explicitly rule out the possibility of their existence in nature. For a long time, the field has remained purely theoretical, being explored through different approaches such as their thermodynamic signatures \cite{Stoilova_2020}, under first-quantized theories \cite{Toppan_2025, Toppan_2021_1, Toppan_2021_2, Balbino_2024}, paramanifold formulations \cite{Zhang_2024}, and as potential candidates for dark matter \cite{Nelson_2016, Kitabayashi_2018}. To date, the only experimental efforts to studying paraparticles are trapped ion simulations of para-Bose and para-Fermi oscillators \cite{Huerta_Alderete_2017_1, Huerta_Alderete_2017_2, Huerta_Alderete_2018, Huerta_Alderete_2021, Huerta_Alderete_2025}.

Recently, an alternative paraparticle formulation was proposed that allows the construction of these systems as emergent quasiparticles \cite{Wang_2025}. This recently introduced $R$\textit{-paraparticle} framework showed that quasiparticle excitations with parastatistical behavior can be constructed using nonlocal string operators from an underlying set of local spin operators. This approach evades the no-go theorems \cite{DHR_I, DHR_II} which assume that excitations are created by local operators. Moreover, the superselection rules \cite{Greenberg_1965} are inherently satisfied by quasiparticle excitations. $R$ paraparticles are predicted to possess novel topological properties in two-dimensional and three-dimensional systems, leading to potential practical applications in quantum communication \cite{wang_2025_2, wang2025_3}.

This then motivates the realization of emergent $R$ paraparticles, potentially in condensed matter systems or trapped ion simulations \cite{Monroe2021}. An indirect method of doing this is to look for exotic many-body signatures in the form of collective excitations and emergent phases that are due to the unconventional exchange statistics of the system. It is therefore interesting to ask how interactions would influence the behavior of $R$ paraparticles. For this purpose, we then consider an $R$ paraparticle system described by an interacting Luttinger model (LM) \cite{Luttinger_1963, Haldane_1981, Voit_1995, schulz_1998}. Conventionally, this describes a one-dimensional (1D) fermionic gas at low temperatures. In this work, we are mainly interested in its property known as bosonization, where fermionic degrees of freedom collectively behave as bosonic modes. This allows the system’s Hamiltonian to be rewritten purely in terms of bosonic operators, making the originally interacting fermionic system exactly solvable \cite{Mahan_2000}. As a result, the model exhibits several non-Fermi liquid properties, including spin–charge separation \cite{Solyom_1979, Vianez_2022}.

In this paper, we study a generalized version of the LM where we extend conventional fermionic operators to $R$-paraparticle operators. We show that bosonization also occurs for $R$ paraparticles with a Fermi-like average occupation number, which we call \textit{$R$ parafermions} in analogy to Green's paraparticles \cite{Green_1953, Greenberg_1965, Stoilova_2020}. This presents a method in which interacting $R$-paraparticle systems can be probed using standard methods of solving bosonic systems \cite{Mahan_2000}. However, before we proceed to the remainder of the study, we first clarify that the term \lq\lq parafermion\rq\rq\ should not be confused with generalizations of Majorana fermions \cite{Fendley_2012, Fendley_2014, Alicea_2016}, which unfortunately share the same terminology in condensed matter literature. More importantly, while the work in \cite{Schmidt_2020} also addresses bosonization of parafermions, our work is completely unrelated to the latter.

The paper shall be organized as follows. In Sec.~\ref{sec:paraparticles}, we review the $R$-paraparticle formulation and clarify how $R$ parabosons and $R$ parafermions are classified based on their thermodynamics and exclusion statistics [cf. Fig.~\ref{fig:1}(a)]. We show that $R$ parafermions obey Pauli's exclusion principle or one of its generalizations, which allows us to define $p$ orderedness in the $R$-paraparticle formalism. In Sec.~\ref{sec:bosonization}, we show that all $R$-parafermionic density wave excitations have bosonic properties independent of the number of degrees of freedom of the $R$ paraparticle, and that the density wave operator always commutes with flavor wave operators, which generalizes the spin wave operator. However, for systems with two internal degrees of freedom, flavor waves are bosonic only for specific $R$-parafermion species, hence only shows flavor-charge separation in specific cases. More importantly, we show that bosonization only applies at low energies for $R$ paraparticles, independent of its exclusion principle, as shown in Fig.~\ref{fig:1}(b). We also show that by introducing interaction, the flavor and density waves of bosonizable $R$-parafermion systems obtain different group velocities. In Sec.~\ref{sec:experiment}, we make use of these ideas to lay out a possible experimental design for realizations of $R$-paraparticle signatures in 1D systems by searching for signs of flavor-charge separation \cite{Vianez_2022} in the absence of spinon excitations. For an ideal underlying system which hosts $R$-paraparticle quasiparticles, we discuss in detail spinor Tonks-Girardeau (TG) gases \cite{Girardeau_1960} and its spin-charge separation behavior, which we summarize in Fig.~\ref{fig:1}(c). Briefly, the boson system possesses a dispersion $\epsilon_k = -2t \cos (k)$ that is purely due to its charges, since spin-charge separation in TG gases implies $v_S = 0$ \cite{Fuchs_2005}. $R$-parafermionic excitations can arise from the spinor TG gas system, and in the LM limit, results in both nonzero flavor- and charge-wave velocities. In the presence of interaction between $R$ paraparticles, $\epsilon_k$ is altered but $v_S$ remains zero, i.e., the spins remain stationary. However for the $R$ parafermions, the flavor and charge velocities become unequal, $v_C \neq v_F$, which becomes a telltale signature of flavor-charge separation. Finally, in Sec.~\ref{sec:conclusion}, we summarize the contents of this paper and provide possible extensions of this work.


\section{$R$ paraparticles}\label{sec:paraparticles}


\subsection{Formulation}

The $R$-paraparticle formulation introduced in \cite{Wang_2025} works as follows. Let $R^{ab}_{cd}$ ($1\leq a,b,c,d \leq m \allowbreak \in \mathbb{Z}_+$) be a four-tensor that satisfies
\begin{subequations}
\begin{align}\label{eq:1a}
    \sum_{c,d} R^{ab}_{cd} R^{cd}_{ef} &= \delta_{ae} \delta_{bf}, \\
    \sum_{g,h,i} R^{ab}_{gh} R^{hc}_{if}R^{gi}_{de} &= \sum_{g',h',i'} R^{bc}_{g'h'} R^{ag'}_{di'} R^{i'h'}_{ef},
\end{align}
\end{subequations}
where the latter is the constant Yang-Baxter equation \cite{Turaev_1988, Majid_1990, etingof_1998}. These conditions impose that an $R$-paraparticle wavefunction returns to itself after two consecutive swaps of the same pair of particles. An $R$ paraparticle that is described by $R^{ab}_{cd}$ then satisfies the generalized commutation relations (GCR)
\begin{subequations}
\begin{align}
    \hat{\psi}_{i,a} \hat{\psi}_{j,b}^{+} &= \sum_{cd} R^{ac}_{bd} \, \hat{\psi}_{j,c}^{+} \hat{\psi}_{i, d} + \delta_{ab} \delta_{ij}, \\
    \hat{\psi}_{i,a}^{+} \hat{\psi}_{j,b}^{+} &= \sum_{cd} R^{cd}_{ab} \, \hat{\psi}_{j,c}^{+} \hat{\psi}_{i, d}^{+}, \\
    \hat{\psi}_{i,a} \hat{\psi}_{j,b} &= \sum_{cd} R^{ba}_{dc} \, \hat{\psi}_{j,c} \hat{\psi}_{i, d},
\end{align}
\end{subequations}
where $\hat{\psi}_{i,a}^+$ ($\hat{\psi}_{i,a}$) creates (annihilates) a particle at mode $i$ with an internal or flavor degrees of freedom indexed by $a$. The integer $m$ denotes the maximum number of internal degrees of freedom considered in the interaction. The creation and annihilation operators satisfy $\hat{\psi}_{i,a}^+ = (\hat{\psi}_{i,a})^{\dagger}$ only if $R$ satisfies
\begin{equation}\label{eq:3}
    \sum_{a,b} R^{ab}_{cd} (R^{ab}_{ef})^* = \delta_{ce} \delta_{df} \quad \text{or} \quad R^{ef}_{ab} = (R^{ab}_{ef})^*.
\end{equation}
The above unitarity requirement imposes a simplified construction of the orthonormal state space due to $R$. When $R$ is nonunitary [i.e., it does not satisfy Eq.~\eqref{eq:3}], an alternate description of the Hermitian inner product is required to impose Hermiticity of obervables \cite{Wang_2025}.

For arbitrary $R$, a contracted bilinear operator
\begin{equation}\label{eq:4}
    \hat{e}_{ij} = \sum_{a=1}^m \hat{\psi}_{i,a}^+ \hat{\psi}_{j,a}
\end{equation}
can be constructed, which obeys
\begin{subequations}\label{eq:5}
\begin{align}
    [\hat{e}_{ij}, \hat{\psi}_{k,b}^+] =& \delta_{jk} \hat{\psi}^+_{i,b},\\
    [\hat{e}_{ij}, \hat{\psi}_{k,b}] =& -\delta_{ik} \hat{\psi}_{j,b},\\
    [\hat{e}_{ij}, \hat{e}_{kl}] =& \delta_{jk} \hat{e}_{il} - \delta_{il} \hat{e}_{kj}.
\end{align}
\end{subequations}
These commutation relations will be crucial in the construction of the bosonization formulation in the subsequent section.

The single-mode partition function for an $R$ paraparticle can be defined as
\begin{equation}\label{eq:6}
    z_R(x) = \sum_{n=0}^{\infty} d_n x^n,
\end{equation}
where $x \equiv e^{-\beta \epsilon}$, $\beta = 1/k_BT$, and $\epsilon$ is the $R$-paraparticle energy. $d_n$ are nonnegative integers and describe the dimension of the $n$-particle state space. Importantly, $d_n = 0$ implies that $n$ $R$ paraparticles cannot occupy the same mode at once. The average occupation number can then be calculated as
\begin{equation}\label{eq:7}
    \langle n\rangle_R = -\frac{1}{\beta} \frac{\partial z_R(x)}{\partial \epsilon} = \frac{x \partial_x z_R(x)}{z_R(x)}.
\end{equation}
In general, different sets of GCRs can result in the same set of $\{ d_n \}_{n=1}^{\infty}$, hence different $R$ paraparticles can possess similar occupation statistics.

For the remainder of this paper, we shall reshape the $R$ tensor as an $m^2 \times m^2$ matrix $M_{uv} = R^{ab}_{cd}$, where $u=m(a-1)+b$ and $v = m(c-1)+d$. As an example, $M$ matrices for the $m=2$ case can be written as
\begin{equation}
    M = \begin{pmatrix}
        R^{11}_{11} & R^{11}_{12} & R^{11}_{21} & R^{11}_{22} \\
        R^{12}_{11} & R^{12}_{12} & R^{12}_{21} & R^{12}_{22} \\
        R^{21}_{11} & R^{21}_{12} & R^{21}_{21} & R^{21}_{22} \\
        R^{22}_{11} & R^{22}_{12} & R^{22}_{21} & R^{22}_{22}
    \end{pmatrix}.
\end{equation}
This makes the elements of $R$ easier to visualize and allows for a convenient expression of complex $R$ tensors. In this $M$-matrix formalism, the requirement Eq.~\eqref{eq:1a} can be rewritten as
\begin{subequations}
\begin{equation}
    M^2 = \mathbbm{1}_{m^2},
\end{equation}
while the unitary requirement Eq.~\eqref{eq:3} becomes
\begin{equation}
    M = M^{\dagger}.
\end{equation}
\end{subequations}


\subsection{$R$ parafermions}\label{subsec:parafermions}

Consider the case where the exclusion statistics due to $R^{ab}_{cd}$ allows at most $n'$ particles to occupy a single mode. This implies that $d_{n>n'} = 0$, hence the average occupation number is reduced to
\begin{equation}
    \langle n\rangle_R = \frac{\sum_{n=1}^{n'} nd_n x^n}{\sum_{n=0}^{n'} d_nx^n}.
\end{equation}
In the low temperature limit of $\beta \to \infty$,
\begin{subequations}
\begin{equation}
        x  \to \begin{cases}
        \infty, \, &\epsilon < 0 \\ 0, &\epsilon > 0
        \end{cases},
\end{equation}
hence
\begin{equation}\label{eq:10b}
    \langle \tilde{n} \rangle_R = \begin{cases}
            n', \, &\epsilon < 0 \\ 0, & \epsilon > 0
        \end{cases} = n'\theta(-\epsilon),
\end{equation}
\end{subequations}
i.e., the occupation number close to absolute zero is proportional to a Heaviside step function $\theta(x)$, and therefore the $R$ paraparticle possesses a Fermi-surface-like structure. This motivates the classification of $R$ paraparticles that satisfy Eq.~\eqref{eq:10b} as $R$ parafermions. Analogously, we show in the Appendix that when $d_{n} \neq 0$ for all $n$, the average occupation of the particle resembles Bose statistics, and therefore these $R$ paraparticles are categorized as $R$ parabosons. A similar classification scheme was also used in Refs.~\cite{li_2025_PRD, li_2025_JHEP}, albeit not expounded in terms of the $R$ paraparticles' occupation statistics. In Fig.~\ref{fig:1}(a), we present several examples of the occupation statistics of $R$-paraparticles described by different single-mode partition functions. The partition functions $z(x) = (1 + x)^m$ are representative of $m$-component fermions, while $z(x) = 1 + mx$ describes $n'=1$ $R$ parafermions. $R$-paraboson occupations described by the dashed lines are obtained from $R'= -R$, where $R$ describes the $R$ parafermion designated with the same curve color. These occupation behaviors were also predicted in Green's paraparticle formulation \cite{Stoilova_2020}.

The requirement that $d_{n>n'} = 0$ implies an exclusion principle is at play within the $R$ paraparticle's GCR. In general, this can be interpreted as the operator algebra satisfying 
\begin{equation}\label{eq:12}
    (\hat{\psi}_{i,a}^+)^{p} \neq 0 \qquad \text{while} \qquad (\hat{\psi}_{i,a}^+)^{p+1} = 0,
\end{equation}
where $p=1$ reduces to Pauli's exclusion principle. This definition is analogous to Green's $p$-ordered parafermions, to which there can be at most $p$ particles in the same quantum state \cite{Greenberg_1965}. We therefore define $p$-ordered $R$ parafermions as $R$ paraparticles that satisfy Eq.~\eqref{eq:12}, i.e., allowing at most $p$ particles of the \textit{same} flavor in a single mode. By definition, we have $n' \geq p$.

For the $m=2$ case, examples of $p=1$ $R$ paraparticles include those that are described by the $M$-matrices
\begin{subequations}
    \begin{align}
        M_1 &= -\mathbbm{1}_4, \label{eq:13a} \\
        M_2 &= \begin{pmatrix}
            -1 & 0 & 0 & 0 \\ 0 & 0 & \gamma & 0 \\ 0 & 1/\gamma & 0 & 0 \\ 0 & 0 & 0 & -1
        \end{pmatrix}, \label{eq:13b}
        \end{align}
and
    \begin{align}\label{eq:13c}
        M_3 &= \begin{pmatrix}
        -1 & 1 & -1 & 0 \\ 0 & -1 & 0 & 0 \\ 0 & -2 & 1 & 0 \\ 0 & 1 & -1 & -1
    \end{pmatrix},
    \end{align}
where $\gamma \in \mathbb{C}_{\ne 0}$. The $R$ tensor defined by $M_1$ gives $n' = 1$ and is exactly Example 3 of \cite{Wang_2025}, while both $M_2$ and $M_3$ have $n' = 2$. Note that the matrices $M_2$ with $|\gamma| \neq 1$ and $M_3$ describe nonunitary $R$ tensors, and ordinary fermions are retrieved from $\gamma = -1$ in Eq.~\eqref{eq:13b}. Likewise, $p=2$ cases also exist for $R$ paraparticles with $m=2$, such as those described by
\begin{equation} \label{eq:13d}
    M_4 = \begin{pmatrix}
            0 & 0 & 0 & \delta \\ 0 & -1 & 0 & 0 \\ 0 & 0 & -1 & 0 \\ 1/\delta & 0 & 0 & 0
        \end{pmatrix},
\end{equation}
\end{subequations}
also with $\delta \in \mathbb{C}_{\ne 0}$. This also describes a nonunitary $R$ tensor whenever $|\delta| \neq 1$. The exclusion statistics of $R$ paraparticles described by $M_4$ go as $\hat{\psi}_{i,a}^+ \hat{\psi}_{i,b}^+ = -\hat{\psi}_{i,a}^+ \hat{\psi}_{i,b}^+ = 0$ and $\hat{\psi}_{i,a}^+ \hat{\psi}_{i,a}^+ \propto \hat{\psi}_{i,b}^+ \hat{\psi}_{i,b}^+ \neq 0$ for $a, b \in \{1,2\}$, $a\neq b$. This results to $\hat{\psi}_{i,a}^+ \hat{\psi}_{i,a}^+\hat{\psi}_{i,a}^+ \propto \hat{\psi}_{i,a}^+\hat{\psi}_{i,b}^+ \hat{\psi}_{i,b}^+ \allowbreak = 0$, implying that three particles cannot occupy the same mode at once, and therefore $d_{n>2} = 0$.

We end this section with an important note that the definition of order $p$ is not an invariant for $R$ paraparticles. This is because under internal basis transformations, one can map an $R$-paraparticle GCR due to $R^{ab}_{cd}$ to another GCR described by $(R')^{ab}_{cd}$, thus following a different set of exclusion statistics. In more detail, we have
\begin{subequations}
    \begin{equation}
    (R')^{ab}_{cd} = S^{ab}_{ef} R^{ef}_{gh} (S^{-1})^{gh}_{cd},
    \end{equation}
    or in terms of the $M$-matrix,
    \begin{equation}\label{eq:3.10b}
    M' \tilde{S} = \tilde{S} M,
    \end{equation}
    where
    \begin{equation}
    S^{ab}_{ef} = Q^{a}_e Q^b_f \qquad \text{and} \qquad \tilde{S} = Q \otimes Q,
    \end{equation}
\end{subequations}
with $Q$ an $m \times m$ invertible matrix. Curiously, under this rule, one can show that $M_2(\gamma = 1)$ is equivalent to $M_4(\delta \in \mathbb{C}_{\neq 0})$, i.e., an $R$ paraparticle of order $p=1$ can get mapped to $p=2$. While we have insufficient information on which internal basis an $R$ paraparticle prefers, the important thing to consider in bosonization is the existence of Fermi-point-like structures in $R$ parafermions, as we show in the next section.


\section{$R$ parafermion bosonization}\label{sec:bosonization}

Having defined $p$ ordering and the classification of $R$-paraparticles into $R$ parafermions or $R$ parabosons, let us now consider the bosonization of 1D interacting $R$ paraparticles. We define the $R$-paraparticle LM kinetic Hamiltonian term as
\begin{subequations}\label{eq:14}
\begin{align}
    \hat{H}_{0} &= \sum_{r,k,a} v_F(rk-k_F) :\hat{\psi}^+_{r,k,a} \hat{\psi}_{r,k,a}: \\
    &= \sum_{r,k} v_F(rk-k_F) \Bigl( \hat{e}_{r, k,k} - n'\theta(k_F - rk) \Bigr) \label{eq:15b}
\end{align}
\end{subequations}
where the constant $\theta(rk - k_F)$ imposes the normal ordering denoted by $:\dots:$, and $\pm k_F$ are the parafermi points of the 1D system. The indices $k$ and $a$ denote momentum and flavor, while $r \in \{+,-\}$ (a mode index) differentiates between the two branches of the dispersion relation $\varepsilon_r(k) = v_F(rk-k_F)$, with $v_F$ as the Fermi velocity. In the LM, the total density operator can be decomposed in terms of the left-moving and right-moving $R$ parafermions, $\hat{\rho}(k) = \hat{\rho}_+(k) + \hat{\rho}_-(k)$, where
\begin{subequations}
\begin{align}
    \hat{\rho}_r(q) &= \sum_{k,a} \Bigl(\hat{\psi}^+_{r,k + q, a} \hat{\psi}_{r,k, a} - \delta_{q,0} \theta(k_F - rk) \Bigr) \\
    &= \sum_k \hat{e}_{r,k+q, q} - n'\delta_{q,0} \sum_{k} \theta(k_F - rk)
    \end{align}
are the density operators for each branch. In a similar manner, flavor wave operators are defined as
    \begin{align}
        \hat{\sigma}(q) &= \hat{\sigma}_+(q) + \hat{\sigma}_-(q), \\
        \hat{\sigma}_r(q) &= \sum_{k,a} \alpha_a \Bigl(\hat{\psi}^+_{r,k + q, a} \hat{\psi}_{r,k, a} - \delta_{q,0} \theta(k_F - rk) \Bigl).\label{eq:15d}
    \end{align}
\end{subequations}
For a direct comparison with the spin-$\frac{1}{2}$ LM, we restrict our results to the case of $m=2$ and define $\alpha_1 = -1$ and $\alpha_2 = +1$ when dealing with flavor waves. Note also that Eq.~\eqref{eq:15d} cannot be expressed in terms of contracted bilinear operators defined in Eq.~\eqref{eq:4} due to the prefactor $\alpha_a$.


\subsection{Density-wave bosonization and flavor-charge separation}

Writing the density operator in terms of $\hat{e}_{r,i,j}$ allows us to evaluate its commutation relation using Eq.~\eqref{eq:5}, resulting in
\begin{equation}\label{eq:16}
    [\hat{\rho}_{r}(-q), \hat{\rho}_{r'}(q')] = \delta_{rr'} \delta_{qq'}\sum_k (\hat{e}_{r,k-q, k-q} - \hat{e}_{r,k, k}).
\end{equation}
We then note that for $R$ parafermions, we can exploit their Fermi-surface structures which leads to the closed-form expression \cite{schulz_1998}
\begin{equation}
    [\hat{\rho}_{r}(-q), \hat{\rho}_{r'}(q')] = r \delta_{rr'} \delta_{qq'} \frac{n'qL}{2\pi},
\end{equation}
($L$ being the system length) dependent only on the maximum number of particles that can occupy a single mode. Comparison with the Bose commutation relation $[\hat{b}_k, \hat{b}^+_k] = \delta_{kk'}$ then motivates the mapping
\begin{equation}\label{eq:19()}
\begin{aligned}
    \hat{b}_q &= \sqrt{\frac{2\pi}{n'qL}} \hat{\rho}_+(-q), \qquad \hat{b}_q^+ = \sqrt{\frac{2\pi}{n'qL}} \hat{\rho}_+(q), \\
    \hat{b}_{-q} &= \sqrt{\frac{2\pi}{n'qL}} \hat{\rho}_-(q), \qquad \hat{b}_{-q}^+ = \sqrt{\frac{2\pi}{n'qL}} \hat{\rho}_-(-q).
\end{aligned}
\end{equation}
Note that because Eq.~\eqref{eq:5} applies to all $R$ tensors, the above equation implies that all $R$-parafermionic LMs possess density waves with bosonic properties. This includes $R$-parafermion types with more than two internal degrees of freedom, and is a property irrespective of order.

Using the same commutation relations of the contracted bilinear operators, it can also be shown that
\begin{equation}\label{eq:19}
    [\hat{\rho}_{r}(-q), \hat{\sigma}_{r'}(q')] = 0,
\end{equation}
which is applicable for all $R$ parafermions with $m=2$. However, to interpret this commutation relation as flavor-charge separation, we must also show that the flavor-wave operator satisfies Bose statistics.


\subsection{Flavor-wave bosonization}

While Eq.~\eqref{eq:19} shows that $\hat{\rho}_r(q)$ and $\hat{\sigma}_r(q)$ always commute for $m=2$ $R$ parafermions, explicit calculation of the commutation relations of the flavor-wave operators shows that it is not always bosonic in nature, hence, Eq.~\eqref{eq:19} does not fully capture flavor-charge separation in that sense. Because we cannot express $\hat{\sigma}_r(q)$ in terms of $\hat{e}_{ij}$, we are required to evaluate the commutator explicitly as
\begin{widetext}
    \begin{align}\label{eq:20}
        \relax [\hat{\sigma}_r(-q), \hat{\sigma}_{r'}(q')] &= \sum_{k,k',a,a'} \alpha_{a} \alpha_{a'} [\hat{\psi}^+_{r,k - q,a} \hat{\psi}_{r,k,a}, \hat{\psi}^+_{r',k' + q',a'} \hat{\psi}_{r',k',a'}] \nonumber \\
        &= \sum_{k,k',a,a'} \alpha_{a} \alpha_{a'} \biggl\{ \delta_{r,r'} \delta_{k-q, k'} \delta_{a,a'} \hat{\psi}^+_{r,k - q,a} \hat{\psi}_{r',k',a'} - \hat{\psi}^+_{r',k' + q',a'} \hat{\psi}_{r',k',a'} \hat{\psi}^+_{r,k - q,a} \hat{\psi}_{r,k,a} \nonumber \\
        &\qquad  + \biggl( \sum_{b,c,d,e,f,g} R^{ab}_{a'c} R^{de}_{ab} R^{a'c}_{ gf} \hat{\psi}^+_{r',k' + q',d} \hat{\psi}^+_{r,k - q,e} \hat{\psi}_{r',k',f} \hat{\psi}_{r,k,g} \biggr) \biggr\}
    \end{align}
showing us that the system bosonizes only if the right-hand side of Eq.~\eqref{eq:20} simplifies to an expression similar to Eq~\eqref{eq:16}. This can only be achieved whenever the last term of the above expression reduces to
\begin{equation}
    \begin{aligned}
        \sum_{b,c,d,e,f,g} R^{ab}_{a'c} R^{de}_{ab} R^{a'c}_{ gf} \hat{\psi}^+_{r',k' + q',d} \hat{\psi}^+_{r,k - q,e} \hat{\psi}_{r',k',f} \hat{\psi}_{r,k,g} = &-\delta_{rr'}\delta_{aa'} \delta_{k-q, k'} \hat{\psi}^+_{r',k'+q',a'} \hat{\psi}_{r,k,a} \\
        &+\hat{\psi}^+_{r',k' + q',a'} \hat{\psi}_{r',k',a'} \hat{\psi}^+_{r,k - q,a} \hat{\psi}_{r,k,a},
    \end{aligned}
\end{equation}
\end{widetext}
or more simply, when the $R$ tensors satisfy
\begin{equation}\label{eq:21}
    \sum_{c,d} R^{ac}_{bd} R^{ef}_{ac} R^{bd}_{hg} = \delta_{e,b} \delta_{h,a} R^{ef}_{hg}.
\end{equation}
This allows for the reduction of the commutator to a closed form
\begin{equation}
    [\hat{\sigma}_{r}(-q), \hat{\sigma}_{r'}(q')] = r \delta_{rr'} \delta_{qq'} \frac{n'qL}{2\pi},
\end{equation}
and thus the bosonic transformation
\begin{equation}
\begin{aligned}
    \hat{c}_q &= \sqrt{\frac{2\pi}{n'qL}} \hat{\sigma}_+(-q), \qquad \hat{c}_q^+ = \sqrt{\frac{2\pi}{n'qL}} \hat{\sigma}_+(q), \\
    \hat{c}_{-q} &= \sqrt{\frac{2\pi}{n'qL}} \hat{\sigma}_-(q), \qquad \hat{c}_{-q}^+ = \sqrt{\frac{2\pi}{n'qL}} \hat{\sigma}_-(-q),
\end{aligned}
\end{equation}
allows for the establishment of $[\hat{c}_k, \hat{c}^+_k] = \delta_{kk'}$.

Returning to our earlier $M$-matrix examples, $M_1$, $M_2(\gamma \in \mathbb{C}_{\neq 0})$, and $M_4(\delta^2 = 1)$ satisfy Eq.~\eqref{eq:21} and hence must have bosonic flavor waves. However, $M_3$ and $M_4(\delta^2 \neq 1)$ do not follow this requirement and therefore do not possess bosonic flavor-wave profiles. We note that even though $M_2(\gamma = 1)$ can be mapped to $M_4(\delta \in \mathbb{C}_{\neq 0})$ under internal basis transformations, the latter case can only satisfy Eq.~\eqref{eq:21} when $\delta = \pm 1$. This shows that Eq.~\eqref{eq:21} is a sufficient but not necessary condition for flavor-wave bosonization, as it can still be satisfied by certain $R$ parafermions after some suitable internal basis transformation. The existence of $R$ parafermions that do not satisfy Eq.~\eqref{eq:21} for any of its internal bases remain an open possibility.


\subsection{Equivalence spectrum}

We now check wether the system described using $R$-parafermion operators is the same as that described by bosonic ones. For $m=2$ $R$ parafermions, it can be shown that the free LM Hamiltonian obeys the commutation relation
\begin{equation}
    [\hat{\phi}_r(q), \hat{H}_0] = v_F r q \hat{\phi}_r(q)
\end{equation}
for $\hat{\phi}_r(q) \in \{ \hat{\rho}_r(q), \hat{\sigma}_r(q) \}$. This allows us to rewrite the kinetic energy Hamiltonian in terms of number operators \cite{Haldane_1981, Voit_1995},
\begin{equation}\label{eq:25}
    \hat{H}_0 = v_F \Bigl[ \sum_{k} |k| \hat{n}^{(b)}_{k} + \frac{\pi}{2L} \sum_{r,a} (\hat{N}_{r,a})^2 \Bigr]
\end{equation}
where $\hat{n}^{(b)}_{k} = \hat{b}_k^+ \hat{b}_k$ and $\hat{N}_{r,a} = \sum_k (\hat{\psi}_{r,k,a}^+ \hat{\psi}_{r,k,a} - \theta(k_F - rk))$. The second summation term takes care of the normal ordering imposed in the $R$-parafermion operator basis. The flavor-wave contribution to the Hamiltonian is the same as Eq.~\eqref{eq:25}, but with $\hat{n}^{(b)} = \hat{b}_k^+ \hat{b}_k$ replaced by $\hat{n}^{(c)} = \hat{c}_k^+ \hat{c}_k$ \cite{Mahan_2000}. Note that these results are applicable only to $R$ parafermions with bosonic flavor waves. The total Hamiltonian under consideration would then be
\begin{equation}
    \hat{H}_T = \hat{H}_0 + \hat{H}_{FW},
\end{equation}
where $H_0$ is given by Eqs.~\eqref{eq:15b} and \eqref{eq:25} in the $R$-parafermion and boson operator bases, respectively.

\begin{figure*}[t]
    \centering
    \includegraphics[width=\textwidth]{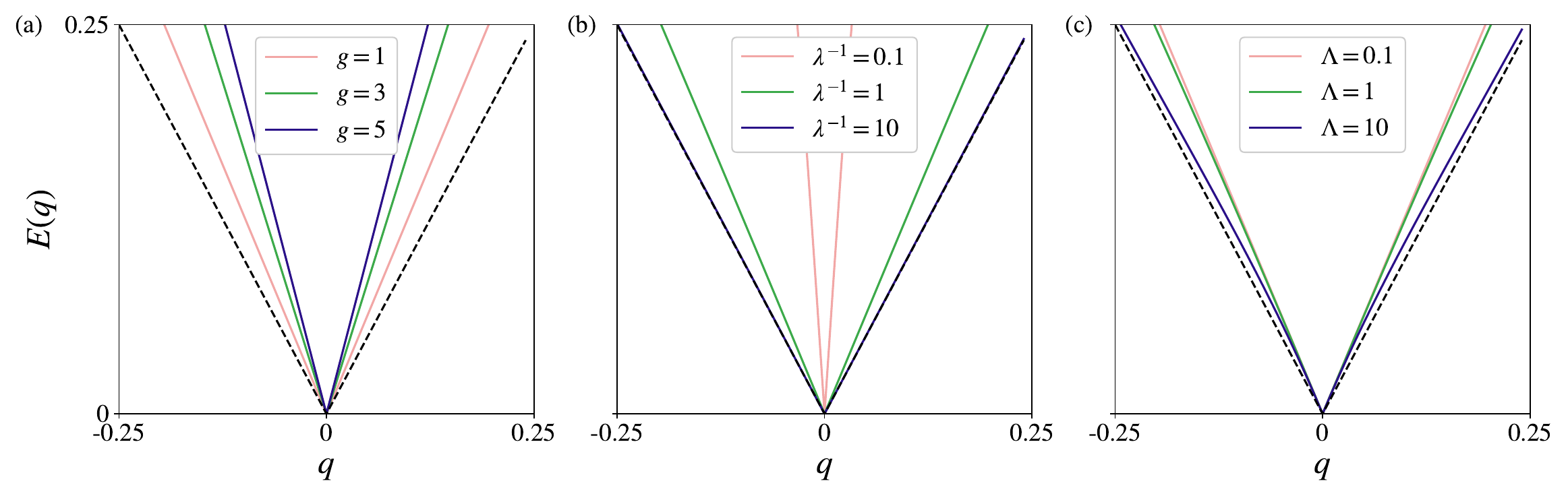}
    \caption{Dispersion relation of $R$-parafermionic LM with (a) contact, (b) Yukawa, and (c) screened interactions, where in (b) and (c), we used $g=1$. We have set $v_F = 1$ for all calculations. For all cases, we see distinct differences between the flavor-wave (denoted by the broken line) and density-wave dispersions (in solid lines). The corresponding density wave velocity is obtained as the slope of the solid curves at $q = 0$.}
    \label{fig:2}
\end{figure*}

One way to verify the correspondence between these Hamiltonian forms is to show that their degeneracies match in both bases. The case where $R$ parafermions have no internal degrees of freedom (i.e., spinless fermions) has already been done in \cite{Haldane_1981} by evaluating the partition functions of each basis separately. Following this method, we make use of the orthonormal $R$-parafermion Fock basis where
\begin{subequations}
\begin{align}
    \hat{e}_{r,k,k} | n_{r,k} \rangle &= \hat{n}_{r, k} | n_{r,k} \rangle = n_{r,k} | n_{r,k} \rangle, \\
    \langle n_{r,k} | n'_{r',k'} \rangle &= \delta_{nn'} \delta_{rr'} \delta_{kk'}.
\end{align}
\end{subequations}
By imposing the latter requirement, we are then restricted to unitary $R$ tensors since nonunitary $R$ requires a more involved inner product definition. The $R$-parafermion partition function is calculated as
\begin{equation}\label{eq:29}
    Z_{\text{PF}}(y) = \prod_{n=1}^{\infty} \biggl( \sum_{j=0}^{n'} y^{2(2n-1)j} \biggr)^4,
\end{equation}
where $y \equiv e^{-\beta v_F \pi/L}$. Note that this expression is independent of the order $p$ of the $R$ parafermion. The $(2n-1)$ exponent is obtained by assuming that the system is under periodic boundary conditions such that $k = 2n\pi/L$ (where $n$ are energy level indices) and $k_F = n_F \pi /L$. Meanwhile, the boson partition function is obtained from the orthonormal basis set $\{ |N_{r,a}, n_{k}^{(b)}, n^{(c)}_{q} \rangle \}$, where
\begin{subequations}
    \begin{align}
    \hat{N}_{r,a} |N_{r,a}, n_{k}^{(b)}, n^{(c)}_{q} \rangle &= N_{r,a} |N_{r,a}, n_{k}^{(b)}, n^{(c)}_{q} \rangle, \\
    \hat{n}^{(b)}_k |N_{r,a}, n_{k}^{(b)}, n^{(c)}_{q} \rangle &= n^{(b)}_k |N_{r,a}, n_{k}^{(b)}, n^{(c)}_{q} \rangle, \\
    \hat{n}^{(c)}_q |N_{r,a}, n_{k}^{(b)}, n^{(c)}_{q} \rangle &= n^{(c)}_q |N_{r,a}, n_{k}^{(b)}, n^{(c)}_{q} \rangle.
    \end{align}
\end{subequations}
These are constructed from the boson vacuum state via
\begin{equation}
|N_{r,a}, n_{k}^{(b)}, n^{(c)}_{q} \rangle = (\hat{U}_{r,a})^{N_{r,a}} \frac{(\hat{b}^+_k)^{n_{k}^{(b)}}}{\sqrt{n_{k}^{(b)}!}} \frac{(\hat{c}^+_q)^{n_{q}^{(c)}}}{\sqrt{n_{q}^{(c)}!}} |0\rangle,
\end{equation}
where $\hat{U}_{r,a}$ is the ladder operator for the $R$-parafermion charge $N_{r,a}$. This comes from a direct generalization of the ladder operators introduced in \cite{Haldane_1981}, and results in the boson partition function
\begin{align}\label{eq:32}
    Z_{\text{B}}(y) &= \biggl[ \sum_{n = -\infty}^{+\infty} (y)^{n^2} \biggr]^{4} \prod_{n=1}^{\infty} \biggl( \frac{1}{1 - y^{4n}} \biggr)^{2} \nonumber \\
    &= \prod_{n=1}^{\infty} \biggl[ \frac{(1 + y^{(2n-1)})^4(1 - y^{2n})^2}{(1 - y^{4n})}\biggr]^2,
\end{align}
where the second equality results from the elliptic theta function equivalence
\begin{equation}
    \theta_3(0,w) = \sum_{m=-\infty}^{+\infty}w^{m^2} = \prod_{n=1}^{\infty} (1+w^{2n-1})^2(1 - w^{2n}).
\end{equation}

Figure~\ref{fig:1}(b) shows a comparison of the partition functions of the LM for both bases. It is interesting to note that $R$ parafermions with different $n'$ have almost equal values of $Z_{\text{PF}}$ even at high temperatures. In contrast to the spinless LM, the boson and $R$-parafermion partition functions do not exactly match. However, $Z_{\text{B}}$ and $Z_{\text{PF}}$ shows a significant overlap in the low-temperature limit, $\beta \to \infty$. This signifies that the spectra described by the Hamiltonians in the boson and $R$-parafermion bases are equivalent, thus implying that bosonization becomes a valid process for 1D systems at low temperatures, generalizing the existing predictions for fermionic LMs.


\subsection{Interacting systems}\label{sec:IIID}

In preparation for the next section, we now consider a more realistic system with interparticle interaction. For simplicity, we make use of the total Hamiltonian
\begin{subequations}
    \begin{align}
        \hat{H}^{\text{LM}} &= \hat{H}_0 + \hat{H}_{\text{FW}} +  \hat{H}_{\text{int}}, \\
        \hat{H}_{\text{int}} &= \frac{1}{2L} \sum_{k} V(k) \hat{\rho}(k) \hat{\rho}(-k). \label{eq:34b}
    \end{align}
\end{subequations}
This can be solved exactly by transforming the Hamiltonian into the boson basis and then applying appropriate diagonalization techniques. The interaction term can be simplified using the mapping given by Eq.~\eqref{eq:19()}, while the free-particle term transforms as Eq.~\eqref{eq:25}. However, under consideration that we conserve the particle number, then we can drop the second term of Eq.~\eqref{eq:25} \cite{Voit_1995}, and the remaining problem becomes equuivalent to a spin-$\frac{1}{2}$ LM, where the total Hamiltonian can be decomposed in terms of density and flavor contributions,
\begin{equation}
    \hat{H} = \sum_{k} E_{\rho}(k) \hat{b}_k^+ \hat{b}_k + E_{\sigma}(k) \hat{c}_k^+ \hat{c}_k,
\end{equation}
with corresponding dispersions \cite{schulz_1998}
\begin{subequations}
    \begin{align}
        E_{\rho}(k) &= v_F|k| \sqrt{1 + \frac{2V(k)}{v_F \pi} }, \\
\text{and} \quad
        E_{\sigma}(k) &= v_F|k|.
    \end{align}
\end{subequations}
In Figs.~\ref{fig:2}(a)-(c), we plot the dispersion relations of interactions of the forms
\begin{subequations}
    \begin{align}
        V(q) &= g, \\
        V(q) &= \frac{g}{q^2 + \lambda^{-2}}, \\
    \text{and} \quad 
        V(q) &= ge^{-\Lambda|q|},
    \end{align}
\end{subequations}
respectively, where the former potential corresponds to a repulsive contact interaction term, while the latter two are medium-ranged interactions, all for $g>0$. This shows that for a simple interacting $R$-parafermionic 1D system described by the LM, one can observe flavor-charge separation as two distinct dispersions with different velocity profiles. Finally, the partition function for the total interacting system is
\begin{equation}
    Z = \prod_{k = -\infty}^{\infty} \frac{1}{(1-e^{-\beta E_{\rho}(k)})(1-e^{-\beta v_F |k|})},
\end{equation}
and thermodynamic variables such as heat capacity and entropy can be calculated after specifying the system's interaction potential.


\section{Potential experimental realization}\label{sec:experiment}

The LM assumes a linearized single-particle dispersion, which makes it valid only in the momentum regime close to $\pm k_F$. This idealization implies that not all real systems possess the necessary properties to exhibit LM properties such as spin-charge separation. Systems that realize these properties are therefore termed Luttinger liquids (LLs). Several works have experimentally verified the existence of LLs by observing some of its key signatures \cite{Bockrath_1999, Gao_2004, Levy_2006, Li_2015, Yang_2017, Vianez_2022}. Specifically in Ref.~\cite{Vianez_2022}, they were able to verify spin-charge separation in a 1D system by looking for two distinct dispersions with different velocity profiles, which were associated with the decoupled spin and charge excitations. Since we showed that $R$-parafermionic LMs also exhibit flavor-charge separation, we should expect that a 1D system containing $R$ parafermions would also display such behavior. 

Following Ref.~\cite{Wang_2025}, suppose that we were able to construct a system of emergent $R$ paraparticles from an underlying 1D fermion system with total particle number $N$. The mapping introduced in the latter ensures that the total particle number is conserved. By writing the Fermi momentum as
\begin{equation}
    k_F = \frac{N\pi}{n'L},
\end{equation}
we can deduce that the validity of the LL theory for the fermion system and the $R$-parafermion quasiparticles overlap if $n'$ are equal for both cases. This may then be problematic, as it becomes difficult to distinguish when the separate dispersions are due to the underlying fermions or to the emergent quasiparticle system. This would be the case for $M_2(\gamma = 1)$ $R$-paraparticle systems in the form of $m$ decoupled chains of $XY$ models which maps to free fermion chains \cite{Chapman_2020}.

In contrast, if the underlying system were bosonic, then we can disregard the issue of Fermi momentum matching and possible dispersion overlaps. Consider for example a 1D spinor Bose gas in the TG limit. In terms of the Bose-Hubbard model, we can write the Hamiltonian as
\begin{equation}
    \hat{H} = -t\sum_{i,a} (\hat{b}_{i,a}^+ \hat{b}_{i+1, a} + \hat{b}_{i+1,a}^+ \hat{b}_{i, a}) - \sum_{i} \mu_i \hat{n}_i,
\end{equation}
where $t$ is the hopping parameter and $\mu_i$ is the lattice chemical potential. The hard-core condition implies $n_i = \{0,1\}$, which imposes the restriction
\begin{equation}
    \hat{b}_{i,a}^+ \hat{b}_{i,a'}^+ = 0, \quad \hat{b}_{i,a}\hat{b}_{i,a'}^+ = \delta_{aa'} \Bigl(1 - \sum_c \hat{b}_{i,c}^+ \hat{b}_{i,c} \Bigr)
\end{equation}
for local operators, while still having $[\hat{b}_{i,a}, \hat{b}_{j,a'}^+] = [\hat{b}_{i,a}, \hat{b}_{j,a'}] = 0$ for $i \neq j$ \cite{Rigol_2004, Rigol_2005}. This allows the bosons to be completely mapped to a system of emergent $R$ paraparticles described by $M_1$, with the Hamiltonian now defined as
\begin{equation}
    \hat{H} = -t \sum_{i,a} (\hat{\psi}_{i,a}^+ \hat{\psi}_{i+1, a} + \hat{\psi}_{i+1,a}^+ \hat{\psi}_{i, a}) - \sum_{i} \mu_i \hat{n}_i.
\end{equation}
Reexpressing the Hamiltonian in momentum basis and expanding the kinetic energy term about $\pm k_F$ while taking $\mu_i = 0$ allows us to obtain an $R$-paraparticle LM with $v_F = 2t \sin k_F$. We can also include interaction terms of the form
\begin{equation}
    \hat{H}_{\text{int}} = \sum_{i \neq j} V_{ij} \hat{n}_i \hat{n}_j
\end{equation}
in the Hamiltonian, which can be mapped in a straightforward manner to $R$-paraparticle operators via
\begin{equation}
    \hat{n}_i = \sum_{a} \hat{b}^+_{i,a} \hat{b}_{i,a} = \sum_{a} \hat{\psi}^+_{i,a} \hat{\psi}_{i,a}.
\end{equation}
This relation can be proven by contracting the flavor indices of the 1D Matrix Product Operator Jordan-Wigner Transformation definition in Ref.~\cite{Wang_2025} (see also its corresponding supplementary information).

The important thing with spinor Bose gases is that while they also exhibit spin-charge separation, the contribution of their spin components to the total energy is small due to a large effective mass in their dispersion. Specifically, in the TG limit, the effective mass becomes infinite and the spins become stationary \cite{Fuchs_2005}. We expect no dispersion profile due to boson spins for an $R$-paraparticle system from an underlying TG gas, and therefore flavor-wave dispersions should be more easily identifiable, as sketched in Fig.~\ref{fig:1}(c). Owing to these properties, spinor TG gases should therefore be ideal candidates for the realization of flavor-charge separation in $R$-parafermion systems.


\section{Summary and Discussion}\label{sec:conclusion}

In this paper, we showed that $R$ paraparticles can be classified as $R$ parafermions or $R$ parabosons based on their occupation statistics. We worked mainly with $R$ parafermions, which have Fermi-surface-like structures and possess generalized exclusion principles. We showed that bosonization applies to certain 1D $R$-parafermion systems described by the LM, in which flavor-charge separation occurs as a result. $R$-parafermionic density waves are shown to have bosonic signatures; however, flavor waves are bosonic in nature only for a subclass of $R$ parafermions. A comparison of the partition function in the LM shows that $R$ parafermions satisfy bosonization only at low temperatures. Nonetheless, these results show promise for potentially observing $R$-paraparticle signatures in 1D systems.

One of the main motivations of this study was to investigate a system of $R$ paraparticles in the presence of interaction. We have shown that for cases where bosonization is applicable, the treatment becomes closely similar to fermionic systems; specifically, their dispersion relations become equivalent. Nonetheless, other generalizations such as the field-theoretic formalism of the LM should not be treated as trivial given the extensive use of generalized commutation relations in their definitions.

A complication to our formalism arises when we consider $R$-paraparticle systems with more than two flavor degrees of freedom, as the current construction of the flavor-wave operator makes sense only for $m=2$, in analogy to the known spin-$\frac{1}{2}$ LM. For a generalized theory, additional operators must be introduced which (i) preserves the number of degrees of freedom in the system under the basis transformation $\{ \hat{\psi}_{r,k,a} \}_{a = 1}^m \to \{\hat{\rho}_r, \hat{\sigma}_{r,1}, \ldots, \hat{\sigma}_{r,m-1} \}$, while at the same time (ii) allows for the description of systems even without prior knowledge of the nature of these degrees of freedom.

Beyond this issue, other possible extensions of this work would be to improve our understanding of the internal basis preference of $R$ paraparticles. Another open problem would be the interpretation of the requirement in Eq.~\eqref{eq:21}, which, due to its role in bosonization for the LM, may be connected to the preferred internal basis of physically realizable $R$ parafermions. If not, then it begs the question of how non-bosonic flavor waves in certain $R$-parafermion systems should be interpreted. Furthermore, it is also interesting to explore the conditions under which emergent $R$ paraparticles appear in systems that are shown to host them.


\section*{Note added}

After publication, we became aware of Ref.~\cite{Sanchez_2024} that appeared in arXiv two months prior to Ref.~\cite{Wang_2025}. The former introduced an alternative classification of quantum statistics beyond ordinary bosons and fermions, termed ``transtatistics'', using a different set of methods rooted in representation theory. Their work predicts several features similar to those of $R$ parastatistics, including, for example, order-one transfermions that exhibit properties closely resembling $M_1$ $R$ parafermions in this work.

\begin{acknowledgments}
We acknowledge M.A. Sulañgi, Z. Wang, and F.N.C. Paraan for their helpful discussions and comments. D.F.S. acknowledges support from the DOST-SEI under the Accelerated Science and Technology Human Resource Development Program.
\end{acknowledgments}

\section*{Data Availability}
There are no publicly available research data or software supporting this manuscript. Requests for further information or data should be sent to the authors.


\appendix* \section{$R$ parabosons}\label{app:a}

In Sec.~\ref{subsec:parafermions}, we discussed the behavior of paraparticles where $d_{n>n'} = 0$ for some finite $n'$. We now discuss what happens in the limit $n' \to \infty$. 

To begin, we first note that the single-mode partition functions of paraparticles satisfy \cite{Wang_2025}
\begin{equation}\label{eq:a1}
    z_R(-x) z_{-R}(x) = 1
\end{equation}
under the premise that $z_R(x)$ is interpreted as the Hilbert series \cite{Polishchuk_2005} of $R$. Suppose $R$ describes an $R$ parafermion, then by Eq.~\eqref{eq:a1},
\begin{equation}\label{eq:a2}
    z_{-R}(x) = \frac{1}{\sum_{n=0}^{n'}d_n(-x)^n} = \sum_{j=0}^{\infty} D_jx^j,
\end{equation}
where the second equality follows the definition of Eq.~\eqref{eq:6}. The coefficients $D_j$ can be obtained by a geometric series expansion of $1/z_R(-x)$ and are expressed as
\begin{equation}\label{eq:a3}
    D_0 = 1, \quad D_j = \sum_{k = 1}^{\min (j,n')} d_k (-)^{k+1} D_{j-k} \in \mathbb{Z}_+.
\end{equation}
However, Eq.~\eqref{eq:a2} cannot be defined in $x \in (0, +\infty)$ as this would result in divergent and negative partition functions in $z_{-R}(x)$. Rather, this definition applies only to $x \in (0, x_0)$, with $x_0 = e^{-\beta \epsilon_0}$ under the following considerations:
\begin{itemize}
    \item Because $d_0 = 1$, i.e., there is only one vacuum state, we require that $1 > |\sum_{n=1}^{n'} d_n (-x)^n|$ for the geometric series expansion to hold. However, because
    \begin{equation}
        z_{-R}(x) = \frac{1}{1 + \sum_{n=1}^{n'}d_n(-x)^n} > 0,
    \end{equation}
    it is automatically guaranteed that $1 > -\sum_{n=1}^{n'} d_n (-x)^n$ is always true for appropriate values of $x$.

    \item Consider otherwise that $x_0 \to \infty$. In this case, we have
    \begin{equation}
        \lim_{x \to \infty} \sum_{n=1}^{n'} d_n (-x)^n = \begin{cases}
        +\infty, \quad \text{for even }n' \\
        -\infty, \quad \text{for odd }n'.
        \end{cases}
    \end{equation}
    However, as implied in the previous bullet, we require $\sum_{n = 1}^{n'} d_n(-x)^n < 1$ to be true within $x\in (0, x_0).$ Clearly, if $x_0 \to \infty$, this condition is violated. Hence, we must impose $x_0$ to be finite, and therefore so must $\beta \epsilon_0$.

    \item At the boundaries of $x \in (0, x_0),$ $z_{-R}(x)$ ceases to be of finite and positive value. However, $z_{-R}(x_0) = 1/z_R(-x_0)\neq 0$ since $z_R(x)$ is a finite degree polynomial with positive finite coefficients, and therefore does not diverge for finite $x$. This therefore implies that
    \begin{equation}
        \lim_{x \to x_0^-} z_{-R}(x) = \lim_{\beta \epsilon \to \beta\epsilon_0^+} z_{-R}(e^{-\beta \epsilon}) = +\infty,
    \end{equation}
    meaning $x_0$ must be the smallest positive real root of $f(x) = z_{R}(-x)$.
    
\end{itemize}
Going back to the average occupation number definition in Eq.~\eqref{eq:7}, we then have
\begin{equation}
    \langle \tilde{n} \rangle_{-R} = \frac{\sum_{n=1}^{n'} nd_n(-)^{n-1}x^n}{\sum_{n=0}^{n'} d_n(-x)^n} = \frac{x\partial_{-x} z_R(-x)}{z_R(-x)}.
\end{equation}
From the earlier discussions, we can therefore rewrite $z_R(-x) = g(-x)(-x+x_0)^{\tau},$ with $\tau$ the multiplicity of the root $x_0,$ and $g(-x)$ a finite degree polynomial possibly containing other real roots larger than $x_0$. Then
\begin{align}
    \langle \tilde{n} \rangle_{-R} &= x\frac{\partial_{-x} z_R(-x)}{z_R(-x)} \nonumber \\
    &= x\frac{g'(-x)(-x+x_0)^{\tau} + \tau g(-x)(-x+x_0)^{\tau-1}}{g(-x)(-x+x_0)^{\tau}} \nonumber \\
    &= \frac{xg'(-x)}{g(-x)} + \frac{\tau x}{(-x+x_0)},
\end{align}
and hence $\langle \tilde{n} \rangle_{-R}$ also diverges at $x_0$. At the same time, because $g(-x)$ has no real roots in $(0, x_0),$ then $\langle \tilde{n} \rangle_{-R}$ is smooth in this region. Finally, one can show that
\begin{subequations}
\begin{align}
    \lim_{\beta \epsilon \to \infty}  \langle \tilde{n} \rangle_{-R} &= 0, \\
    \partial_{\beta \epsilon} \langle \tilde{n} \rangle_{-R} \Bigl|_{\beta \epsilon \to \infty} &\propto -d_1,
\end{align}
\end{subequations}
hence within the domain $\epsilon \in (\epsilon_0, \infty)$, the average occupation number is exponentially decreasing from infinity, reminiscent of Bose statistics as in Fig.~\ref{fig:1}(a). This motivates us to classify paraparticles that have single-mode partition functions in the form of Eq.~\eqref{eq:a2} with $D_n \neq 0$ for all $n$ as $R$ parabosons.

While the earlier discussion shows that all $R$ parafermions with a four-tensor $R$ have corresponding $R$ parabosons described by $-R$, the converse is not always true. Consider the examples
\begin{subequations}
    \begin{align}
        M_5 &= \begin{pmatrix}
            \zeta & 0 & 0 & 0 \\ 0 & 0 & \kappa & 0 \\ 0 & \kappa & 0 & 0 \\ \eta & 0 & 0 & -\zeta 
        \end{pmatrix}, \\
        M_6 &= \begin{pmatrix}
            \zeta & 0 & 0 & \eta \\ 0 & 0 & \kappa & 0 \\ 0 & \kappa & 0 & 0 \\ 0 & 0 & 0 & -\zeta
        \end{pmatrix},
    \end{align}
and
    \begin{align}
        M_7 &= \begin{pmatrix}
            \zeta & 0 & 0 & 0 \\ 0 & 0 & \delta & 0 \\ 0 & 1/\delta & 0 & 0 \\ 0 & 0 & 0 & -\zeta
        \end{pmatrix},
    \end{align}
\end{subequations}
where $\zeta, \kappa = \pm 1, \, \eta \in \mathbb{C}, \, \delta \in \mathbb{C}_{\ne 0}$. In all cases, no exclusion principle can be extracted from their GCRs, and therefore all possible combinations of $\delta, \, \zeta, \, \eta,$ and $\kappa$ would result in $d_{n} \neq 0$ for all $n$. Their single-mode partition functions and average occupation number are therefore fully described by
\begin{subequations}
\begin{align}
	z(x) &= \sum_{n=0}^{\infty} d_n x^n, \\
    \langle \tilde{n} \rangle &= \frac{xz'(x)}{z(x)}= \frac{\sum_{n=1}^{\infty} n d_n x^n}{\sum_{n=0}^{\infty} d_n x^n}.
\end{align}
\end{subequations}
However, because $d_n \in \mathbb{Z}_+$, $z(x)$ can only admit positive real values up to some finite $x' \in (0, 1]$ such that
\begin{subequations}
    \begin{equation}
    \lim_{x \to (x')^-} z(x) = \lim_{x \to x'} \biggl( \sum_{n=0}^{\infty} d_n x^n \biggr) = +\infty.
    \end{equation}
    This also results in
    \begin{equation}
    \lim_{x \to (x')^-} z'(x) = \lim_{x \to x'} \biggl( \sum_{n=1}^{\infty} nd_n x^{n-1} \biggr) = +\infty,
    \end{equation}
\end{subequations}
but because the power series structure of $z'(x)$ grows faster than $z(x)$ for increasing index $n$, then
\begin{equation}
\lim_{x \to (x')^-} \frac{z'(x)}{z(x)} = + \infty,
\end{equation}
and therefore $\langle \tilde{n} \rangle$ also diverges at $x'$. Combined with the fact that
\begin{equation}
\lim_{\beta \epsilon \to +\infty} \langle \tilde{n} \rangle = \lim_{x \to 0} \frac{xz'(x)}{z(x)} = 0,
\end{equation}
and with $z(x)$ and $z'(x)$ being continuous and smooth between $x \in (0, x')$, then for $x' = e^{-\beta \epsilon_0}$, $\langle n \rangle_R$ should still be under the $R$-paraboson classification as in Fig.~\ref{fig:1}(a). Remarkably, this implies that for $R$ paraparticles, there are more $R$ parabosons than $R$ parafermions modulo internal basis transformations.



%

\end{document}